# Microstructure evolution and densification during spark plasma sintering of nanocrystalline W-5wt.%Ta alloy

Ajeet K. Srivastav[1*], Suresh Bandi[1], Abhishek Kumar[2] and B.S. Murty[3,4]

[1]Department of Metallurgical and Materials Engineering, Visvesvaraya National Institute of Technology, Nagpur - 440010, India

[2]Institute for Frontier Materials, Deakin University, Geelong, Geelong, VIC 3216, Australia

[3]Indian Institute of Technology Hyderabad, Kandi, Sangareddy - 502285, India

[4]Department of Metallurgical and Materials Engineering, Indian Institute of Technology Madras, Chennai - 600036, India

**Abstract**

The present work reports the effect of Ta on densification and microstructure evolution during non-isothermal and spark plasma sintering of nanocrystalline W. Nanocrystalline W-5wt.%Ta alloy powder was synthesised using mechanical alloying. The nanocrystalline powder was characterised thoroughly using X-ray diffraction line profile analysis. Furthermore, the shrinkage behaviour of nanocrystalline powder was investigated during non-isothermal sintering using dilatometry. Subsequently, the alloy powder was consolidated using spark plasma sintering up to 1600°C. The role of Ta on stabilising the microstructure during spark plasma sintering of nanocrystalline W was investigated in detail using electron backscatter diffraction. The average grain size of spark plasma sintered W-5wt.%Ta alloy was observed as 1.73±1 µm.

**Keywords:** Nanocrystalline alloys; non-isothermal sintering; Kirkendall porosity; spark plasma sintering

---

[*] Corresponding Author

Email: srivastav.ajeet.kumar@gmail.com

# 1. Introduction

Tungsten (W) has evolved as a primary candidate for plasma-facing material in nuclear fusion devices owing to its high melting point [1,2], low sputtering coefficient [3], high thermal conductivity [4], and low radioactive fuel retention [5,6]. However, the inherent low-temperature brittleness limits its use for structural applications [7,8], particularly as this property severely worsens under irradiation conditions [9,10]. Also, low fracture toughness, high-temperature oxidation, and recrystallisation embrittlement are other drawbacks to be considered [11]. In addition to W-based composite materials, alloying and imparting nanocrystallinity in the W have been suggested for improving the ductility and bringing down the ductile to brittle transition temperature (DBTT) in the material [12–14]. Besides, the grain refinement and refinement-induced large grain-boundary volume fractions are encouraging for oxidation and radiation resistance [15,16]. However, alloying could be a possible approach, along with grain refinement, to overcome the drawbacks mentioned above by improving the material's properties.

Alloying W with tantalum (Ta) brings the possible enhancement in the ductility, decreases the water oxidation, helps in mitigating surface cracking caused by the high-fluence threshold for nanostructure formation on the surface of W [5,17]. In addition, Ta lowers the cluster formation under irradiation condition and hence the chances of embrittlement in W and W alloys [18]. However, W-based binary alloys are known for Kirkendall porosity which makes them difficult to consolidate conventionally [19,20]. Therefore, for such alloy powders it is crucial to choose an appropriate sintering process for improving the grain refinement and the densification of the final structure. Recently, spark plasma sintering (SPS) has been established as a promising consolidation technique for various materials. The process can facilitate a high degree of densification at the lowest temperatures and for shorter periods with better grain refinement [21,22]. Moreover, Srivastav *et al.* [23] has shown that localised diffusion-couple-induced pore evolution could be advantageous for densifying such materials by adopting the SPS process on account of the inherent localised heating mechanisms.

In the current work, a nanocrystalline W-5wt.%Ta alloy powder was synthesised by mechanical alloying. First, the microstructure evolution and shrinkage behaviour of the alloy powder was studied during non-isothermal sintering using dilatometry. Subsequently, the alloy powder was consolidated via SPS to achieve a nearly 95% relative density with better grain refinement. The study demonstrates that SPS can be a potential consolidation route to densify such materials compared to conventional methods.

**2. Experimental details**

The W-5wt.%Ta alloy was prepared via mechanical alloying using a high-energy planetary ball mill (Fritsch Pulversitte-5, Germany) equipped with a set of tungsten carbide (WC) vials and balls (10 mm diameter) for grinding. The milling was employed for 6 h at 300 rpm using a 10:1 ball-to-powder ratio with toluene as a process-controlling agent.

X-ray diffraction profile of mechanically milled W-5wt.%Ta alloy powder was collected using an X'Pert Pro (PANalytical, The Netherlands) X-ray diffractometer with Cu-K$\alpha$ radiation in the 2$\theta$ range 10–140° using 0.02° step size with a 30 s time per step. A line-profile analysis [23–26] was used for profile fitting using the pseudo-Voigt function in High Score Plus software (version 3.0e). Furthermore, the full width at half-maximum (FWHM) and the shape parameter ($\eta$) from the fitted profile were used as input parameters for the double-Voigt analysis [27,28] using the BREADTH programme (version 4) [29] for the quantification of crystallite size and root-mean-square strain (RMSS) in the milled powder sample. It is worth noting here that the FWHM and $\eta$ input parameters are transformed to the Cauchy $(\beta_C)$ and the Gaussian $(\beta_G)$ part of the corresponding Voigt function using the approximations as discussed earlier [30,31].

Dilatometry studies were performed on a cold compact (a cylindrical shaped pellet with 8 mm diameter) of milled powder using a dilatometer, Setaram, Setsys Evolution. The cold compact was sintered non-isothermally in an argon atmosphere with a constant heating rate of 20 K min$^{-1}$ up to 1600°C under ~ 0.6 N load. The effect of alloying with Ta on the densification behaviour of nanocrystalline W was explored with the support of shrinkage data obtained during the experiment.

Consolidation of the W-5wt.%Ta powder was achieved using SPS (spark plasma sintering, Dr. Sinter SPS-650 machine, Sumitomo Metals, Japan). Densification of the powder was carried out in a cylindrical graphite die with 10 mm diameter. The process was taken up to 1600°C with a heating rate of 100°C min$^{-1}$ and a dwell period of 5 min. A uniaxial pressure of 50 MPa was applied throughout the process.

Microstructural investigations and elemental analysis were conducted on non-isothermally sintered and SPS samples using a scanning electron microscope (FEI QUANTA 400 manufactured by FEI, USA) equipped with an energy-dispersive X-ray spectrometer (EDAX–AMETEK Materials Analysis Division, USA). Crystal orientation measurements were made on SPS sample using an HKL Oxford system attached to an FEG-SEM (Zeiss Gemini 300).

## 3. Results and discussion

An XRD pattern of milled W-5wt.%Ta alloy powder is illustrated in Fig.1(a). The magnified (200) peak as shown in the inset clearly indicates the broadened peak profile and undissolved Ta peak. Besides instrumental effects, the physical origin of the peak broadening is attributed to diffraction-order-independent 'coherently diffracting domain (crystallite) size' and diffraction-order-dependent 'lattice distortion/strain' in reciprocal space [32].

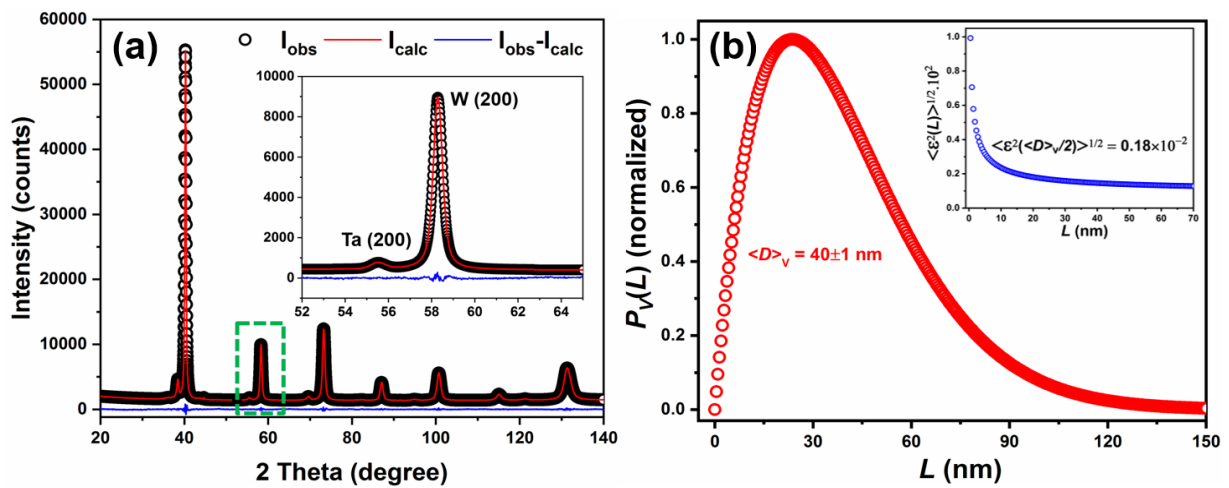

**Fig.1.** (a) The XRD pattern of milled W-5wt.%Ta alloy powder with the fitted profile function, and (b) the crystallite size distribution and the root-mean-square strain over the column length.

Practically, it is always a concern how to deconvolute size and strain effects. Nowadays, the double-Voigt approach has been widely accepted for such deconvolution [32]. Here, a Voigt function approximation for both size- and strain-broadened profiles is considered.

The Fourier transform of the Voigt function in terms of distance, $L$, having size and strain components can be represented as an analytical expression [33,34]:

$$A(L) = \exp\left[-2L\beta_C - \pi L^2 \beta_G^2\right] \qquad (1)$$

The size ($S$) and strain ($D$) contributions can be deconvoluted with the assumption that the Cauchy and the Gaussian functions represent the size and strain parts, respectively [27,35,36]. However, both the size and the strain effects can again be approximated by Voigt functions to tackle the non-ideality situation. As the convolution of two Voigt functions is again a Voigt function, the Cauchy ($\beta_{SC}$ and $\beta_{DC}$) and Gaussian ($\beta_{SG}$ and $\beta_{DG}$) components for both the size and strain can be represented as [27,34,36]:

$$\beta_C = \beta_{SC} + \beta_{DC} \qquad \text{2(a)}$$

$$\beta_G^2 = \beta_{SG}^2 + \beta_{DG}^2 \qquad \text{2(b)}$$

The $\beta_{SC}$, $\beta_{DC}$, $\beta_{SG}$, and $\beta_{DG}$ unknowns can be determined using input parameters for two/or more peaks of the profile.

Furthermore, the area $\left(P_S(L)\right)$ and the volume-weighted $\left(P_V(L)\right)$ column length distribution functions can be obtained by inserting the $\beta_{SC}$ and $\beta_{SG}$ values in equation 1 to obtain the size coefficient $\left(A_S(L)\right)$ and using the relation [27,34,36]:

$$P_S(L) \propto \frac{d^2 A_S(L)}{dL^2}; \quad P_V(L) \propto L \frac{d^2 A_S(L)}{dL^2} \qquad (3)$$

From equation 3, the surface-weighted ($<D>_S$) and volume-weighted ($<D>_V$) column length can be calculated using the expression [27,34,36]:

$$<D>_{S,V} = \frac{\int_0^\infty L P_{S,V}(L) dL}{\int_0^\infty P_{S,V}(L) dL} \qquad (4)$$

Also, the mean-square strain (MSS) as a function of averaging distance $L$ can be obtained by using the expression [27,34,36]:

$$<\varepsilon^2(L)> = \frac{1}{s^2}\left[\frac{\beta_{DG}}{2\pi} + \frac{\beta_{DC}}{\pi^2}\left(\frac{1}{L}\right)\right] \qquad (5)$$

The structural parameters of the milled powder were obtained by following the double-Voigt approach as discussed above. Typical for any distribution function, the Cauchy component ($\beta_{SC}$, $2.33\times10^{-3}$ Å$^{-1}$) of the size-broadening profile dominates over the Gaussian component ($\beta_{SG}$, $0.612\times10^{-3}$ Å$^{-1}$). However, the Gaussian component ($\beta_{DG}$, $1.11\times10^{-3}$ Å$^{-1}$) of the strain-broadening profile overshadows the Cauchy component ($\beta_{DC}$, $0.898\times10^{-3}$ Å$^{-1}$). This indicates that the strain contribution throughout the domains is predominated over by either the fluctuation in the density of embedded atoms or dislocation-cell walls rather than the strain around the individual dislocations [37–39]. It should be pointed out here that the integral breadth was contributed to more by size than by the strain part. Therefore, strain does not play a major role in terms of peak broadening in the present case [27].

The crystallite size (column length) distribution for W is shown in Fig.1(b). The volume-weighted average crystallite size was found to be as $40\pm1$ nm. The inset shows the root-mean-square strain (RMSS) as a function of $L$. Typical of the double-Voigt analysis, the RMSS falls off with $1/L$. The RMSS averaged over a distance $<D>_v/2$ was estimated as $0.18\times10^{-2} \pm 0.2\times10^{-4}$. The lattice parameter for W in the mechanically milled W-5wt.%Ta was calculated using the Nelson-Riley method [40]. The lattice parameter and the Nelson-Riley function (NRF) corresponding to each peak reflection were calculated. The precision lattice parameter was obtained from the intercept of a linear fitting of lattice parameter versus NRF plot.

The calculated lattice parameter was 0.31644 nm, which is a decrease of 0.013% and 0.025% when compared to mechanically milled W with the same milling parameters and unmilled W powder, respectively [25]. The reduced crystallite size and interfacial stresses evolved during the milling could reduce the lattice parameter of the milled W. However, the decrement in lattice parameter is doubled for W in the presence of Ta in the present case compared to an earlier report for milled W [25]. In fact, Ta possesses a higher atomic radius and a higher lattice parameter than W. In principle, the lattice parameter of W should increase with alloying by Ta. As reported by Romig and Cieslak [41], the interdiffusion distances between W and Ta were found to be <1μm at 1300°C after 220 days. This additionally confirms the extremely sluggish diffusion kinetics at the W/Ta interface. The Ta evidenced in the XRD pattern indicates that the Ta has not shown considerable tendency to diffuse at the W/Ta interface. In addition, it has been shown that W diffuses 100 times faster in pure Ta than does Ta in W [42]. Consequently, W has more of a tendency to dissolve in Ta than Ta does in W. Also, the Ta lattice parameter was estimated as 0.31914 nm, which is lower than that of the as-received Ta powder (0.33011 nm). Hence, Ta becomes a Ta-rich solid solution in the course of milling and further processing [43]. In addition, grain-boundary interfacial stresses accompanied by the evolution of reduced crystallite size and non-equilibrium grain boundaries [26], should result in the lower lattice parameter of W as calculated and discussed above.

Fig.2(a) shows the shrinkage (%) and shrinkage rate (%/min) during non-isothermal sintering of the nanocrystalline W-5wt.%Ta alloy powder. There is first an expansion up to 10.5% at 1000°C followed by a shrinkage up to 5.6% at the highest temperature of 1600°C.

Initially, the maximum expansion rate was observed as 1.25% /min at approximately 700°C. Later, the maximum shrinkage rate was 1.1% /min at nearly 1400°C. The pronounced expansion can be well understood in view of a Kirkendall effect arising from the different atomic mobilities of W and Ta as reported by Lenz and Riley [19]. The Kirkendall porosity might play a major role in the early expansion of W-Ta, followed by a sudden drop as shrinkage caused by enhanced interdiffusion at higher temperatures and aided by alloying where the Kirkendall effect plays its role in a positive way [20,23].

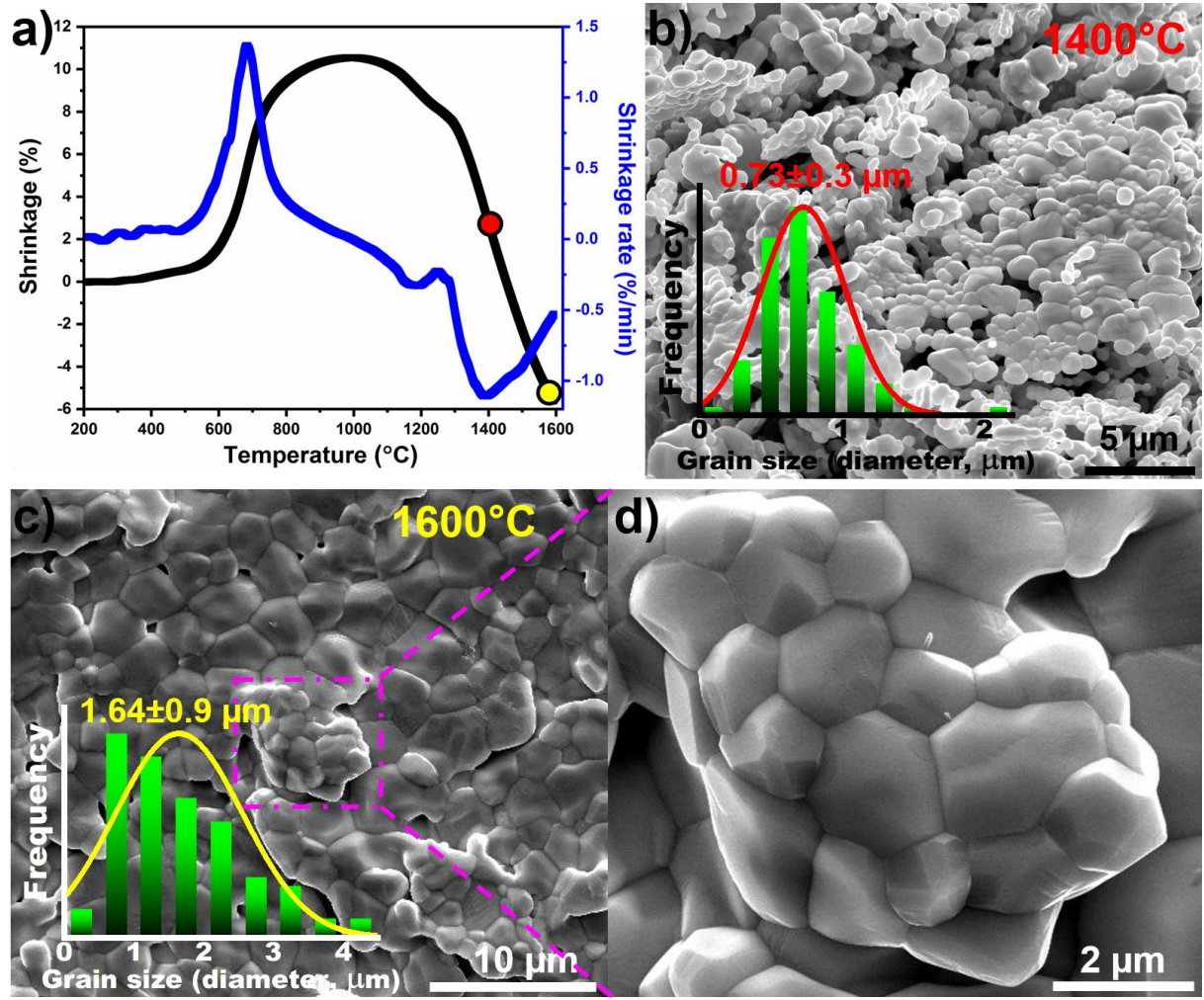

**Fig.2.** (a) The shrinkage and shrinkage rate of nanocrystalline W-5wt.%Ta alloy powder during non-isothermal sintering, and the morphology and the grain-size distribution for the sample non-isothermally sintered up to (b) 1400°C, and (c) 1600°C. Magnified view of the locally evolved faceted and grainy structure from the square section in (c) is illustrated in (d).

To understand this further, the experiment was performed at an intermediate temperature up to 1400°C as well. The morphological evolution as illustrated in Fig.2(b) shows a Kirkendall-effect-induced porous structure with locally interconnected grains. The grain-size distribution is shown as an inset. The average grain size was estimated as 0.73 ± 0.3 µm. The localized densification improves as sintering proceeds further up to 1600 °C as illustrated in Fig.2(c). However, whole sintering does not change anything at the larger scale beyond those occurring in the initial stages of sintering in terms of shrinkage behavior (maximum 5.6%). The grain-size distribution and the average grain size were estimated in the light of the localized densification as illustrated in the inset to Fig.2(c). As expected, the localized densification enhances with grain growth. The average grain size was found to be 1.64±0.9 µm. A magnified view of the evolved grainy structure with well-developed grain boundaries and facets is shown in Fig.2(d).

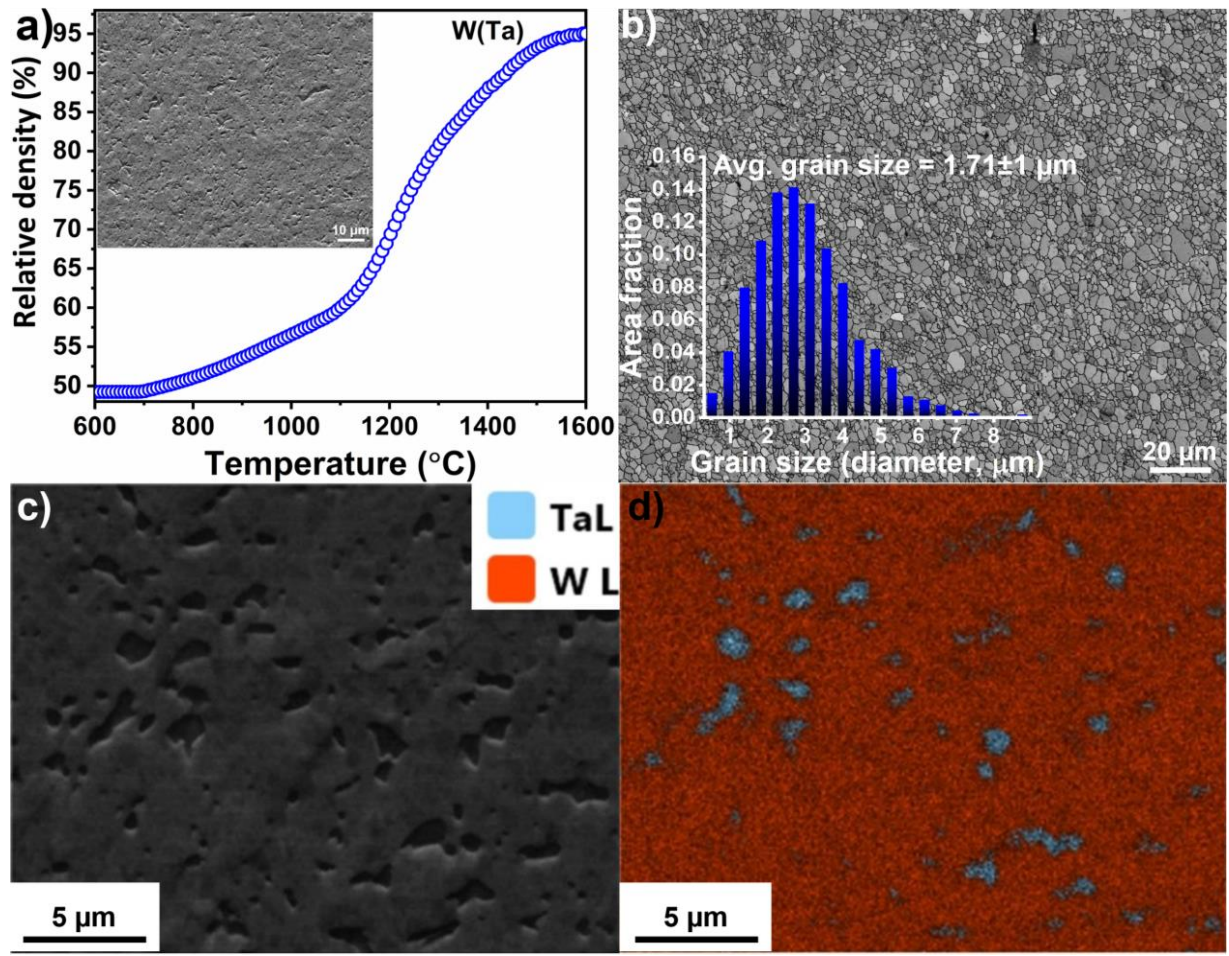

**Fig.3.** (a) The densification curve during SPS of nanocrystalline W-5wt.%Ta alloy powder with secondary electron micrograph (inset), and (b) image quality (IQ) map showing the grain and grain boundaries with grain-size distribution (inset), (c) back-scatter electron miocrograph, and (d) elemental mapping showing the two-phase structure with Ta-rich solid solution and W as matrix.

Unlike non-isothermal sintering, the densification initially occurs at a slower rate followed by a steep rise in density until finally saturating at nearly 95% relative density. The SEM micrograph, shown as an inset, confirms the dense microstructure after the SPS. The microstructure of the consolidated sample was investigated using EBSD. An image quality (IQ) map was constructed from the EBSD data for the SPS consolidated sample. The obtained IQ map, shown in Fig.3(b), clearly shows the grains and grain boundaries that evolved during the sintering process. The grains are mostly uniform in size, with an average size estimated as 1.73 ± 1 µm. Interestingly, the average grain size of the SPS consolidated sample is close to the grain size as observed during non-isothermal sintering up to 1600°C. Almost similar grain sizes, irrespective of the densification process, suggests the same origin of microstructural stability during the sintering process. It is worth noting here that the pure W processed under similar conditions results in grain sizes almost 6.5 times bigger than in the present case [23].

To understand the role of Ta on grain-growth retardation during SPS of nanocrystalline W, elemental mapping was performed on the sample. The elemental analysis of the microstructure shows a two-phase structure with a Ta-rich phase along with the W as the matrix phase. The darker contrast in the back-scatter SEM image is mapped as Ta in light-blue colour in the elemental mapping, whereas W, showing a greyish colour in the back-scatter SEM image, is mapped in orange colour. As discussed above, the Ta becomes a Ta-rich solid solution [43] appearing as dark contrast in the BSE image and confirming with the elemental mapping. The thermodynamic model prediction by Murdoch and Schuh [44] reports a weak segregation tendency of Ta in the W-Ta system. Therefore, the retardation in grain-boundary mobility, and hence finer grain sizes of SPS W-5wt.%Ta, could be ascribed to a kinetic stabilization of the microstructure originating from a solute drag effect caused by the Ta.

## 4. Conclusions

In conclusion, the shrinkage/densification behavior and the microstructure evolution during non-isothermal and SPS of nanocrystalline W-5wt.%Ta have been studied. An extensive Kirkendall-pore-induced expansion was observed during the non-isothermal sintering of the alloy. Almost 95% relative density was achieved during SPS. The densification during SPS could be ascribed to dynamic pore-evolution-mediated enhanced point contacts which further help in localized heating. Interestingly, the average grain size for non-isothermally sintered (1.64 ± 0.9 µm) and the SPS W-5wt.% Ta alloy (1.73 ± 1 µm) were found to be similar. Finally, it has been established that SPS could be a potential consolidation route to produce nano/ultrafine grain W-Ta alloys.

**Acknowledgements**

We are grateful to Dr. Jyoti Shankar Jha for his extended support in characterization of the samples.

**Disclosure statement**

No potential conflict of interest was reported by the author(s).

**Funding**

Ajeet K. Srivastav gratefully acknowledges the financial support from CSIR-India (via award no. 09/084(0519)2010-EMR-I).